\documentclass[prb,reprint,twocolumn,superscriptaddress,showpacs]{revtex4-1}
\usepackage{amsmath}
\usepackage{amssymb}
\usepackage{natbib}
\usepackage{epstopdf}
\usepackage{graphicx}
\usepackage{color}
\usepackage{times}
\usepackage[colorlinks,bookmarks=false,citecolor=blue,linkcolor=red,urlcolor=blue]{hyperref}

\def\be {\begin{equation}}
\def\ee {\end{equation}}

\begin{document}
\title{Ground-State Energy and Spin Gap of Spin-$1/2$ Kagom\'e Heisenberg Antiferromagnetic Clusters: Large Scale Exact Diagonalization Results}
\author{Andreas M. L\"auchli}
\email{aml@pks.mpg.de} \affiliation{Max-Planck-Institut f\"ur Physik komplexer Systeme, N\"othnitzer Str.\ 38, D-01187 Dresden, Germany}
\affiliation{Institut f\"ur Theoretische Physik, Universit\"at Innsbruck, A-6020 Innsbruck, Austria}
\author{Julien Sudan}
\affiliation{Institut Romand de Recherche Num\'erique en Physique des Mat\'eriaux (IRRMA), CH-1015 Lausanne, Switzerland}
\author{Erik S. S{\o}rensen}
\email{sorensen@mcmaster.ca} \affiliation{Department of Physics and Astronomy, McMaster University, Hamilton,
   Canada L8S 4M1}

\begin{abstract}
We present a comprehensive list of ground state energies and spin gaps of finite kagom\'e clusters with up to 42 spins obtained 
using large-scale exact diagonalization techniques. This represents the current limit of this exact approach. For a fixed number
of spins $N$ we study several cluster shapes under periodic boundary conditions in both directions resulting in a toroidal geometry.
The clusters are characterized by their side length and diagonal as well as the shortest
"Manhattan" diameter of the torii.  
A finite-size scaling analysis of the ground state energy as well as the spin gap is then performed
in terms of the shortest toroidal diameter as well as the shortest "Manhattan" diameter. The structure of the spin-spin correlations 
further supports the importance of short loops wrapping around the torii.
\end{abstract}
\maketitle

The quest for magnetic materials and model systems exhibiting quantum spin liquid behavior is
of considerable current experimental and theoretical interest. Among such systems, the kagom\'e $S=1/2$ 
antiferromagnet stands out as a prototypical highly frustrated quantum magnet in two spatial 
dimensions~\cite{Balents2010} that potentially could exhibit spin liquid behavior. However,
a complete understanding of this deceptively simple model given by,
\begin{equation}
H=J\sum_{\langle i,j\rangle}{\bf S}_i\cdot{\bf S}_j,
\end{equation}
has proven surprisingly difficult
and exact diagonalization (ED) results on this model, despite their limitation to very modest system sizes,
have become crucial for both theoretical developments and non-exact numerical techniques.

Several Exact Diagonalization (ED) simulations were carried out
to explore puzzling facets of the kagom\'e antiferromagnet~\cite{Elser1989,
Zeng1990,Chalker1992,Leung1993,Elstner1994,Lecheminant1997,Waldtmann1998,Sindzingre2000,
Waldtmann2000,Richter2004a,Lauchli2009,Sorensen2009,Sindzingre2009}. ED 
 established some of its remarkable properties, such as
the absence of magnetic order and the enormously high number of singlet excitations below the lowest
spinful excitation. Complementary computational techniques have also been applied to the kagom\'e 
Heisenberg antiferromagnet including series expansions~\cite{Singh1992,Singh2007,Singh2008}, Quantum Monte 
Carlo~\cite{Nyfeler2008}, Diagonalizations in the nearest-neighbor valence bond basis and variants thereof~\cite{Zeng1995,Mambrini2000,Poilblanc2010}, 
Contractor Renormalization (CORE)~\cite{Budnik2004,Capponi2004}, 
Multi-scale entanglement renormalization ansatz (MERA)~\cite{Evenbly2010}, and the Density Matrix 
Renormalization group (DMRG)~\cite{Jiang2008,Yan2010}.

In this work we report on large scale ED results for the ground state energy and the spin
gap of the Heisenberg $S=1/2$ antiferromagnet on various kagom\'e samples consisting of up to $N=42$ spins. 
This considerable increase in system sizes (and Hilbert space size) was made possible by a 
distributed memory parallelization of our ED codes. We provide a finite size scaling of the
ground state energy and the spin gap as a function of the shortest diameter of the torii, which seems to capture the finite size
dependence in a more systematic way than a simple $1/N$ scaling used previously~\cite{Lecheminant1997,Waldtmann1998,Sindzingre2009}.

\begin{table*}[ht]
\begin{center}
\begin{tabular}{|c|c|c|c|c|c|c|r|r|r|} 
\hline $N$&$ {\bf a}, {\bf b}$ &$|{\bf a}|$&$|{\bf b}|$&$d$& $d_M$&$|G|$ &Total $E$ ($J$) & $E/N$ & $\Delta$ \\ 
\hline
\hline
{\bf 12} &{\bf (2,0), (0,2)}&	{\bf 2}&{\bf 2}&{\bf 2} & {\bf 4} & {\bf 48} & {\bf --5.444 875 216}   &	{\bf --0.453 740}  & {\bf 0.382 668 366   }\\ 
\hline 
15 &(2,-1), (-1,3) &	$\sqrt{3}$&$\sqrt{7}$&$\sqrt{7}$& 4 & 20 & --6.589 143 829   & 	--0.439 276  &  0.418 800 403 \\
\hline
18 a&(2,-1), (0,3) & $\sqrt{3}$&3&$\sqrt{12}$	& 4 &24			     &	 --8.064 482 605   & --0.448 027 & 0.270 115 263   \\
18 b&(2,-2), (-2,-1)& $2$&$\sqrt{7}$&$3$ &  4  &48 	     &	  --8.048 270 773   & --0.447 126 & 0.284 567 177   \\
\hline
21 &(2,1), (-1,3)& $\sqrt{7}$&$\sqrt{7}$&$\sqrt{7}$& 6  & 336 & --9.172 279 619   & --0.436 775  & 0.278 637 026   \\
\hline
24 &(1,2), (-3,2)&$\sqrt{7}$&$\sqrt{7}$&$\sqrt{12}$ & 6 &96 &  --10.589 965 547   & --0.441 249  & 0.207 828 742  \\
\hline
27 a&(2,1), (-3,3) &$\sqrt{7}$&3&$\sqrt{13}$ & 6 & 18  & --11.793 996 213  & --0.436 815   &  0.275 413 255   \\
{\bf 27 b}&{\bf (3,0), (0,3) }&{\bf 3}&{\bf 3}&{\bf 3} & {\bf 6} &{\bf  	216}	 & {\bf --11.779 504 985}  &{\bf   --0.436 278} & {\bf 0.268 776 803}  \\
\hline
30 & (2,1), (-2,4) &$\sqrt{7}$&$\sqrt{12}$&$\sqrt{13}$ & 6 & 20 & --13.154 318 948   & --0.438 477 &  0.152 855 536  \\
\hline
33  & (1,2), (4,-3) &$\sqrt{7}$&$\sqrt{13}$&$\sqrt{19}$  & 6 & 22 & --14.410 195 048   & --0.436 673 & 0.229 455 039  \\
\hline
36 a&(-2,3), (4,0) &      $ \sqrt{7}$& 4&$\sqrt{19}$     & 6 & 24   & --15.787 874 847   & --0.438 552 & 0.144 945 554  \\
36 b&(3,0), (-3,4) &   $3$&$\sqrt{12}$&$\sqrt{21}$         & 6 & 48 & --15.806 927 756  & --0.439 081 & 0.170 275 671    \\
36 c&(3,0), (-1,4) &    $3$&$\sqrt{13}$&$4$                & 6       & 24 & --15.814 334 002  & --0.439 287 & 0.184 874 846   \\
{\bf 36 d}&{\bf (4,-2), (-2,4)} & $\bf \sqrt{12}$&$\bf \sqrt{12}$&$\bf\sqrt{12}$  &{\bf  8} & {\bf 144} & {\bf --15.781 555 118}  & {\bf --0.438 377} & {\bf 0.164 189 901}   \\
\hline
39 a  &(-1,3), (5,-2)& $\sqrt{7}$&$\sqrt{19}$&$\sqrt{21}$ &  6 & 26 &   --17.038 187 797   & --0.436 877  & 0.199 163 545   \\
39 b & (1,3), (-3,4) & $\sqrt{13}$&$\sqrt{13}$&$\sqrt{13}$  & 8 & 78 & --17.020 192 866  & --0.436 415 & 0.222 433 924    \\
\hline
42 a & (-1,3), (5,-1) & $\sqrt{7}$&$\sqrt{21}$&$\sqrt{28}$ & 6 & 28  & --18.395 959 984  &     --0.437 999  & 0.120 425 \phantom{111}  \\
42 b & (-2,4), (4,-1) &$\sqrt{12}$&$\sqrt{13}$&$\sqrt{19}$ & 8 & 28 & --18.401 988 921  &  --0.438 143 & 0.149 092 139  \\
\hline
\end{tabular}
\end{center}
\caption{Cluster studied in this work. Listed are: The number of spins $N$. The basis vectors ${\bf a},{\bf b}$ in terms of ${\bf a_1}$ and ${\bf a_2}$ (each of length $2a$).
The length of the basis vectors, ($|{\bf a}|,|{\bf b}|$), in units of $2a$ along with the length of the diagonal $d=\min(|{\bf a}-{\bf b}|,|{\bf a}+{\bf b}|)$.
The Manhattan length, $d_M$, the  length of the shortest loop wrapping around the
torus.
The number of elements of the symmetry group $|G|$. The total ground-state energy, $E$. The energy per site, $E/N$. The value of the spin gap, 
$\Delta$, between the $S=0$ ground-state and the lowest $S=1$ state for even samples or the gap from the $S=1/2$ ground-state to the lowest $S=3/2$ state for odd samples.
Results shown in bold are for clusters with the full symmetry of the Kagom\'e plane.}
\label{tab:energies}
\end{table*}

\section{The Clusters}
\begin{figure}
\centerline{\includegraphics[angle=90,width=0.9 \linewidth]{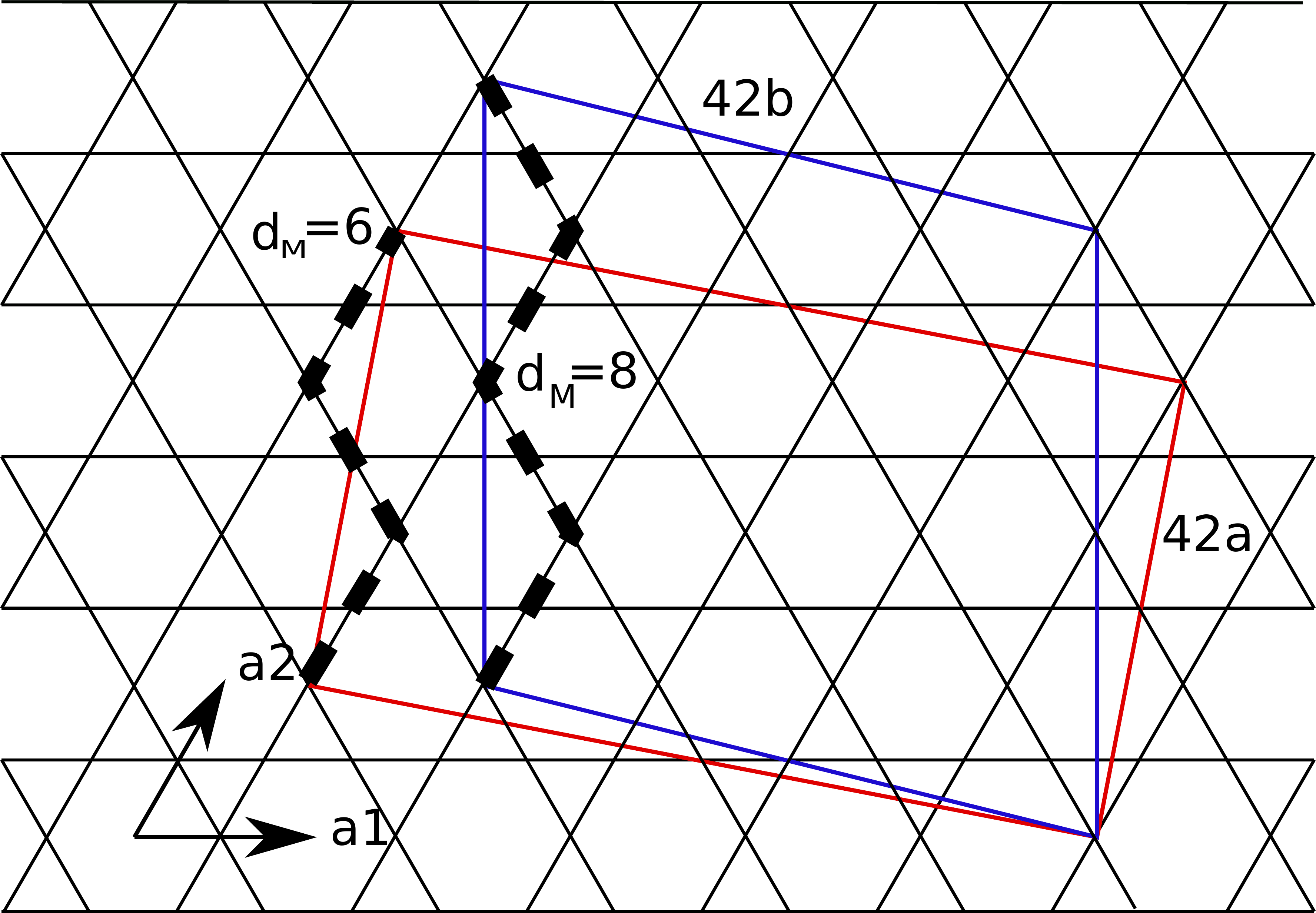}}
\caption{(Color online) The two 42-site kagom\'e clusters used in this study. The shortest Manhattan diameters,
$d_M$, are shown as thick dashed lines.
}
\label{fig:N42}
\end{figure}
The list of the clusters used in this study along with their properties is given in Table~\ref{tab:energies}. For a
given number of spins, $N$, we list in most cases several clusters. For a given $N$ there are many such clusters and
the ones we list are chosen to be close to optimal, where optimal refers to a cluster that would have the length
of both sides as well as the shortest diagonal all equal. This is considered optimal since such a cluster would be more
likely to have the full point group symmetry of the infinite kagom\'e lattice. Of the clusters we consider only 12, 27b
and 36d have the full symmetry of the kagom\'e plane and they are therefore shown in bold. The basis vectors ${\bf a,b}$ of all clusters are given in terms
of the vectors ${\bf a_1, a_2}$ shown in Fig.~\ref{fig:N42} where the two largest clusters 42a and 42b are shown.
We also list the shortest {\it Manhattan diameter}, $d_M$, of each cluster. This measure can be visualized by tiling the plane
with the cluster and finding the shortest path between two equivalent sites walking along the bonds of the lattice. 
$d_M$ is then simply equal to the number of bonds traversed. We also define the shortest {\it geometrical diameter}, this
is simply Min($|a|,|b|$). We also list the number of elements of the symmetry group, $|G|$, for nearest-neighbor interactions
on the cluster. Note that clusters 18b, 21, 24, 27b have larger symmetry groups than expected based on the applicable
symmetries of the infinite kagom\'e lattice.

\section{ED on distributed systems}
Typically an ED study of a cluster would involve a complete symmetry analysis of the cluster and resulting spectra.
One could then write the Hamiltonian matrix restricted to a given symmetry sector to a file that would be read into memory for diagonalization. 
Using {\it shared memory} systems this has been
achieved in the past for a specific quantum number sector for
42 spins on the star lattice~\cite{Richter2004}. However, this approach becomes
impossible for the larger system sizes and symmetry sectors
included in this study, because writing or storing the Hamiltonian
becomes prohibitively slow and matrix elements have
to be calculated on-the-fly. Furthermore, the largest sectors
do not fit into accessible shared memory machines anymore,
such that these large system sizes are {\it only} treatable on {\it distributed
memory} systems where each computational node
only has a relatively small addressable memory space, typically
a few giga-bytes (or less). In general, it is much easier and cheaper
to scale a distributed system to a large combined memory space than a
shared memory system and future large ED studies will likely have to be performed
on distributed systems. Due to the physical constraints of a distributed memory
system some peculiarities remain: A Lanczos diagonalization proceeds by iteratively performing,
$N_{it}$, matrix vector multiplications on vectors of length, $M$. If treating one element
of the vector requires a time $t$, the full cpu-time for the calculation is roughly, $N_{it} M t$. Due to the memory constraints 
of distributed systems the application of a non-trivial symmetry that would reduce $M$ by a factor of $K$ almost always will increase $t$ by
a factor {\it larger} than $K$, even when the number of cpu's used is the same. 
The result is that it is {\it slower} to diagonalize the smaller symmetry reduced Hamiltonian than the full one. For the largest clusters
we therefore only use very few simple symmetries. 
For the 42 site clusters when no lattice symmetries were used the resulting maximal Hilbert space dimension is then
${42 \choose 21}/2=269'128'937'220$.

\section{Results}

\begin{figure}
\centerline{\includegraphics[width=0.9\linewidth]{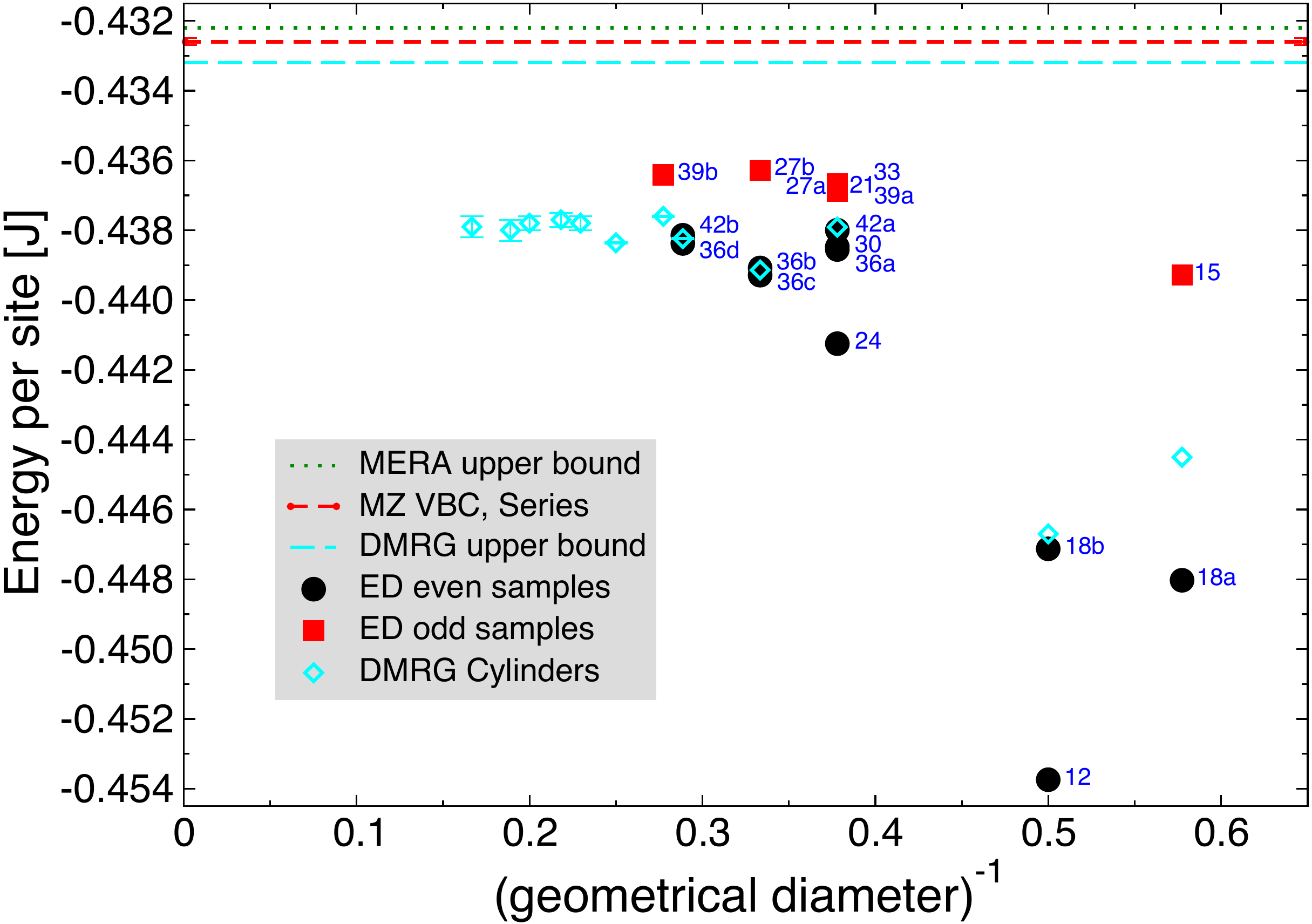}}
\caption{(Color online) Ground state energy per site plotted as a function of the inverse shortest geometrical diameter.
Remarkably good agreement with recent DMRG results~\cite{Yan2010} on long cylinders with the same diameter is observed. For comparison
we also display the MERA upper bound~\cite{Evenbly2010}, series expansion results for the 
Marston-Zeng valence bond crystal~\cite{Singh2007,Singh2008}, as well as the DMRG upper bound~\cite{Yan2010}.}
\label{fig:erg_manhattan}
\end{figure}

\begin{figure}
\centerline{\includegraphics[width=0.9\linewidth]{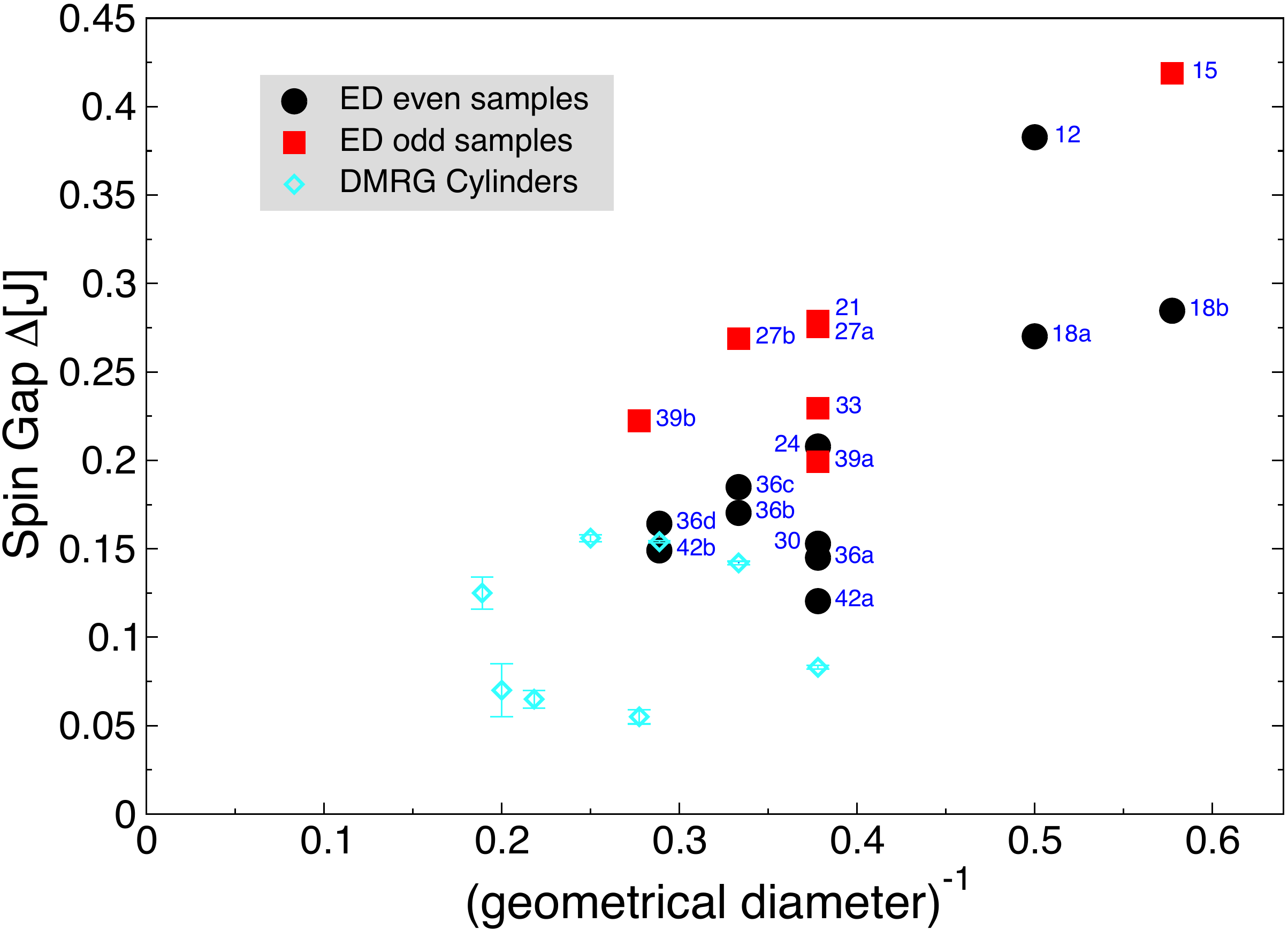}}
\caption{(Color online) Spin gap of even and odd kagom\'e samples obtained by 
ED and plotted as a function of the inverse shortest geometrical diameter. 
For comparison recent DMRG data~\cite{Yan2010} obtained on long cylinders with the same
diameter are shown.}
\label{fig:spingap}
\end{figure}

Our main result is the table \ref{tab:energies} listing the ground state energy and the spin gap of the samples considered
in this work. Some of the smaller samples have already been studied in the past and energies were quoted
for 12,15,18b, and 21 in Ref.~\onlinecite{Zeng1990}~\cite{note1},
for 27b in Refs.~\onlinecite{Leung1993,Lecheminant1997} and for 36d in Refs.~\onlinecite{Leung1993,Waldtmann1998}. 
An approximate ground state energy and spin gap of the sample 36c have been obtained by DMRG in Ref.~\onlinecite{Jiang2008}.

We plot the ground state energies as a function of the inverse geometrical diameter, which in our convention corresponds to
$1/|\mathbf{a}|$ and display the result in Fig.~\ref{fig:erg_manhattan}. This presentation seems to capture the 
finite size effects in a more systematic way than the previously used $1/N$ scaling, as the data seems to behave consistently
upon increasing the system size, while keeping the diameter constant. Furthermore we observe good agreement with recent
DMRG data~\cite{Yan2010} on long cylinders with the same diameter, thus corroborating the accuracy of the DMRG simulations.

Next we plot the spin gap data in Fig.~\ref{fig:spingap} in the same way. For each diameter we observe that the spin gap is monotonously
decreasing with system size (for even and odd samples separately). While it is difficult to extrapolate the ED spin gap for constant diameter due
to a lack of system sizes for large diameters, the qualitative agreement with DMRG results on long cylinders suggests that it is indeed the diameter which
controls the finite size effects upon moving towards the two-dimensional bulk limit.

\begin{figure}
\centerline{\includegraphics[width=\linewidth]{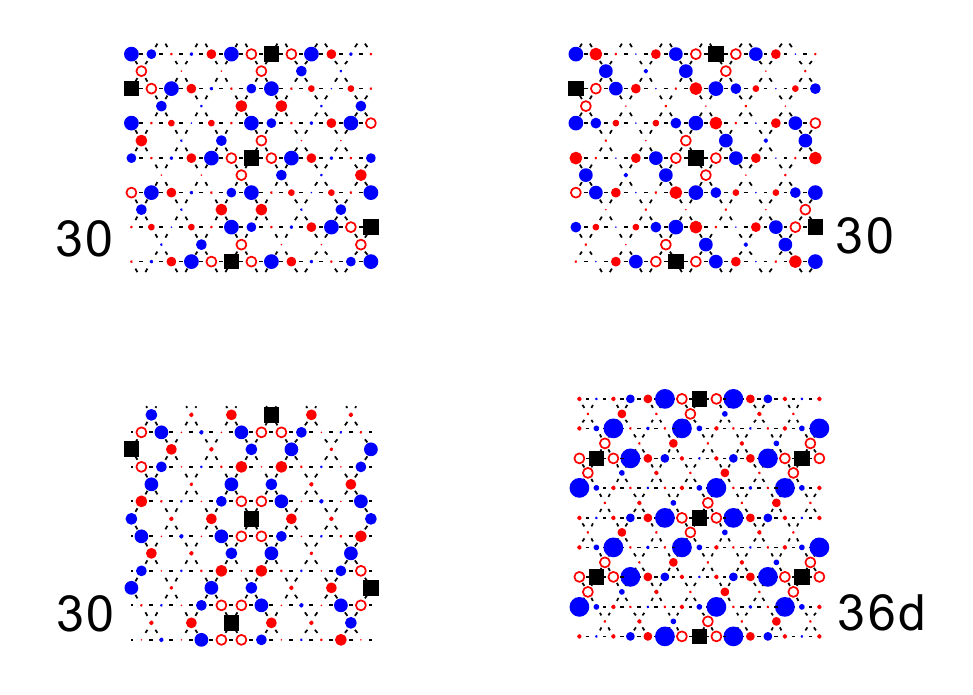}}
\caption{(Color online) Magnified spin-spin correlations in the ground state
of the 30 and 36d kagom\'e samples. The 30 site cluster with its low symmetry requires 
three distinct reference sites to be considered. The diameter of the circle is 
proportional to the magnitude of the spin-spin correlation with the reference site (black filled square).
Blue (red) color denotes positive (negative) correlations.
The nearest neighbor correlations (all antiferromagnetic) have been dropped for clarity. 
One observes pronounced staggered spin-spin correlations along selected loops wrapping 
around the torus in the left two panels and the lower right panel.}
\label{fig:spinspincorrs}
\end{figure}

Complementary evidence for the important role played by short loops wrapping around the torus is provided by a magnified view on the 
spin-spin correlations in the ground state of several even samples displayed in Fig.~\ref{fig:spinspincorrs} 
(see Ref.~\onlinecite{Leung1993} for tabulated values of sample 36d). In a spin liquid with a very short correlation length
one would expect spin correlations between distant sites to be very weak and also not to depend significantly on the sample geometry. Indeed in 
Fig.~\ref{fig:spinspincorrs}, most spin-spin correlations are quite weak, but pronounced staggered spin spin correlations along selected loops 
wrapping around the sample are revealed. We expect these resonances to disappear once the samples are sufficiently wide.

\section{Conclusions}
We have presented a systematic study of the ground-state energies and spin gaps of many Kagom\'e clusters.
In particular we have presented results for $N=39,42$ obtained on distributed memory parallel clusters. Future
quantum numbered resolved ED studies on the largest samples should allow to check the appearance of the 
topological degeneracy required for the $\mathbb{Z}_2$ spin liquid advocated in Ref~\onlinecite{Yan2010}.

\acknowledgments 

We thank D.~A.~Huse for useful correspondence on sample geometries
and are we grateful to D.~A.~Huse, R. Moessner and S.~R.~White for discussions.

AML acknowledges support by the MPG RZG and allocation of computing time on the PKS-AIMS and the BlueGene/P machines at the MPG RZG.
ESS acknowledges allocation of computing time at the Shared Hierarchical Academic Research Computing Network (SHARCNET:www.sharcnet.ca)
and support by NSERC.

\end{document}